\documentclass[a4paper,11pt]{article}
\pdfoutput=1 

\usepackage[no-natbib-sort]{jheppub} 
                     
\usepackage{amssymb,amsmath,amsthm,graphicx}
\usepackage{latexsym}
\usepackage{bm}
\usepackage[]{hyperref}
\usepackage{cleveref}

\def\be{\begin{equation}}
\def\ee{\end{equation}}
\def\ba{\begin{eqnarray}}
\def\ea{\end{eqnarray}}

\def\nn{{\it \nonumber}}

\begin{document}

\title{Higher derivative corrections to Kerr black hole thermodynamics}

\author{Harvey~S.~Reall and}
\author{Jorge~E.~Santos}
\affiliation{Department of Applied Mathematics and Theoretical Physics, University of Cambridge, Wilberforce Road, Cambridge CB3 0WA, UK} 

\emailAdd{hsr1000@cam.ac.uk}
\emailAdd{jss55@cam.ac.uk}

\abstract{
In an effective field theory approach to gravity, the Einstein-Hilbert action is supplemented by higher derivative terms. In the absence of matter, four derivative terms can be eliminated by a field redefinition. We use the Euclidean action to calculate analytically the corrections to thermodynamic quantities of the Kerr solution arising from terms with six or eight derivatives. The eight derivative terms make a non-negative correction to the entropy (at fixed mass and angular momentum) if their coefficients have appropriate signs. The correction from the six derivative terms does not have a definite sign. 
}

\maketitle
\section{Introduction}
Effective field theories are a useful tool for studying physics at low energy without knowing the details of physics at high energy. The idea is to write an action as an infinite expansion of terms with increasing numbers of derivatives, multiplied by powers of some UV length scale $L$. For physics at scales much larger than $L$, only the first few terms in this expansion are required and so one can make predictions once a few coefficients have been determined either by observation or by matching to a UV theory. 

Applying this philosophy to gravity implies that the Einstein-Hilbert action must be supplemented by higher-derivative corrections. If one assumes diffeomorphism invariance then, if the metric is the only relevant light field, these correction terms must be built from the curvature of spacetime. Terms in the action which vanish on-shell can be eliminated by field redefinitions. In four dimensions, this implies that four-derivative terms can be eliminated (or are topological) and so the action takes the form
\be
 I =  \frac{1}{16 \pi G_N} \int d^4 x \sqrt{-g} \left( -2 \Lambda + {\cal R}  +  {\cal L}_6 + {\cal L}_8 + \ldots \right).
 \ee
Henceforth we will set $\Lambda=0$. It has been argued that field redefinitions can be used to bring the $6$-derivative terms to the form \cite{Endlich:2017tqa,Cano:2019ore}
\be
  {\cal L}_6 = \eta_e L^4 R_{ab}{}^{cd} R_{cd}{}^{ef} R_{ef}{}^{ab} + \eta_o L^4 R_{ab}{}^{cd} R_{cd}{}^{ef} \tilde{R}_{ef}{}^{ab} 
 \ee
and the $8$-derivative terms to the form \cite{Endlich:2017tqa}
\be
 {\cal L}_8 = \lambda_e L^6 {\cal C}^2 + \tilde{\lambda}_e L^6 \tilde{\cal C}^2 + \lambda_o L^6 \tilde{\cal C} {\cal C}.
\ee
In the above expressions, $L$ is a length scale below which UV physics becomes important, $\eta_e$, $\eta_o$, $\lambda_e$, $\tilde{\lambda}_e$ and $\lambda_o$ are dimensionless coupling constants and
\be
 \tilde{R}_{abcd} = \epsilon_{ab}{}^{ef} R_{efcd} \qquad {\cal C} = R_{abcd} R^{abcd} \qquad \tilde{\cal C} = R_{abcd} \tilde{R}^{abcd}.
\ee
The subscripts $e$ and $o$ on the coupling constants indicate whether the terms that they multiply are even or odd under orientation reversal (i.e. under parity or time-reversal transformations). 

Usually we assume that the UV scale $L$ is very small, for example the Planck length. But Ref. \cite{Endlich:2017tqa} has emphasized that if we take an agnostic view of UV physics then $L$ could be much larger, perhaps large enough that higher derivative corrections have small but observable effects on astrophysical black holes. For this to happen, $L$ would have to be of the order of a kilometer. One might think this is excluded by laboratory tests of gravity but Ref. \cite{Endlich:2017tqa} argues that this is not the case (provided the UV theory is suitably ``soft") because such tests are performed in a weak-field environment, where the higher derivative terms are suppressed because the curvature is small. 

The above effective action also captures the effects of ``integrating out" massive fields starting from a conventional theory of gravity minimally coupled to matter. If one has a free field of mass $m$ minimally coupled to gravity then loops involving the massive field generate higher derivative terms in the gravitational effective action (see e.g. \cite{Goon:2016mil}). In the above notation this would correspond to the replacements $\eta_i L^4 \rightarrow c_i/(m^2 M_p^2)$ in ${\cal L}_6$ and  $\lambda_i L^6 \rightarrow d_i/(m^4 M_p^2)$ in ${\cal L}_8$ where $M_p = 1/\sqrt{8\pi G_N}$ is the Planck mass and $c_i$ and $d_i$ are dimensionless constants. Such terms are very small. For a black hole of radius $r_+$ one can only integrate out the massive field if $m r_+ \gg 1$ which gives $m \gg 10^{-10} {\rm eV}$ for a kilometer sized hole. But even with $m \sim 10^{-10} {\rm eV}$ the higher derivative terms in the Lagrangian are (at the horizon) $10^{-76}$ times the size of the Einstein-Hilbert term for such a hole. 

Putting aside any possible astrophysical implications, it is interesting to calculate the effect of higher derivative corrections on black hole solutions of GR simply because of the universal nature of these corrections. It is also possible that consistency of low energy black hole physics may lead to constraints on the coefficients of the higher derivative terms, which may supply information about the unkown UV theory. 

In this paper we will study the effect of the higher derivative terms on the Kerr black hole \cite{Kerr:1963ud}, arguably the most important solution of GR. It only makes sense to calculate corrections perturbatively to first order in the various coupling constants because higher orders in perturbation theory will be sensitive to terms in the action with more than eight derivatives. Previous work has studied the even parity ${\cal L}_6$ corrections to the Schwarzschild solution \cite{Lu:1993sq,Dobado:1993id}. More recently, the ${\cal L}_8$ corrections to Schwarzschild and slowly rotating Kerr black holes have been calculated \cite{Cardoso:2018ptl}. Finally, the ${\cal L}_6$ corrections to the Kerr metric have been determined perturbatively as an expansion to high order in $J/M^2$ \cite{Cano:2019ore} where $J$ is the angular momentum and $M$ the mass of the hole. 

Determining higher derivative corrections to the Kerr metric analytically seems very difficult. However, in this paper we will show that one can calculate the corrections to {\it thermodynamic} properties of the black hole {\it without determining the correction to the metric}.\footnote{Having said this, we have also determined (numerically) the correction to the metric \cite{reallsantos}.} We will calculate the first order correction to the Euclidean action of the corrected black hole solution. In principle this correction is the sum of two pieces: (a) the higher derivative terms evaluated on the unperturbed Kerr solution, and (b) the shift in the Einstein-Hilbert action arising from the higher-derivative corrections to the metric. However, we will show that the contribution (b) always vanishes if one works in the grand canonical ensemble, with fixed temperature and angular velocity.  (This is essentially because the Kerr metric extremizes the unperturbed action but in the proof we need to consider surface terms carefully.) Hence the perturbation to the Euclidean action can be calculated explicitly simply by evaluating the higher derivative terms on the Kerr solution.\footnote{It was observed some time ago that a similar result is true for higher-derivative corrections to anti-de Sitter black branes \cite{Gubser:1998nz} or black holes \cite{Caldarelli:1999ar}. This observation was based on determining the correction to the action via an explicit calculation of the correction to the metric. As far as we are aware, no {\it a priori} explanation for this observation has been given previously.} From this we can obtain all thermodynamic quantities.

This calculation shows that the parity-odd terms in ${\cal L}_6$ and ${\cal L}_8$ do not affect thermodynamic quantities (to first order) so we focus on the even terms. The fact that the even terms in ${\cal L}_8$ have a definite sign implies that they give a correction to the (Wald) entropy at fixed $M,J$ that also has a definite sign. These corrections to the entropy are non-negative if, and only if, $\lambda_e \ge 0$ and $\tilde{\lambda}_e \ge 0$. Interestingly, these are exactly the same bounds as obtained from an analysis of unitarity and analyticity of graviton scattering amplitudes \cite{Bellazzini:2015cra}, and from the absence of superluminal graviton propagation \cite{Gruzinov:2006ie}. The significance of this is unclear because the even term in ${\cal L}_6$ gives an entropy correction whose sign depends on $J/M^2$. We will discuss this further at the end of the paper. 

As a taster for our results, we will now present the results for extremal black holes. In the extremal limit we find that the mass, entropy and angular velocity of the black hole are given in terms of the angular momentum as\footnote{We choose units so that $G_N=1$ henceforth.}
\begin{subequations}
\begin{align}
& M = |J|^{1/2} + \frac{1}{14}\frac{L^4}{|J|^{3/2}}\eta_e -\frac{L^6}{|J|^{5/2}}\left(\frac{76}{5}\lambda_e + \frac{296}{5}\tilde{\lambda}_e\right)  + \ldots \label{eq:Mext}
\\
& S = 2\pi |J| -\frac{2\pi}{7} \frac{L^4}{|J|}\eta_e-\frac{L^6}{J^2}\frac{\pi}{5}\left[\frac{1}{4}\left(4864+1575 \pi\right)\lambda_e+\left(4736+1575 \pi\right)\tilde{\lambda}_e\right]  + \ldots \label{eq:Sext}
\\
& \Omega_H = \mathrm{sgn\;}J\;\left\{ \frac{1}{2\,|J|^{1/2}}-\frac{3}{28}\frac{L^4}{|J|^{5/2}}\eta_e+\frac{L^6}{|J|^{7/2}}\left(38\,\lambda_e+148\,\tilde{\lambda}_e\right)+\ldots\right\} \,.\label{eq:angext}
\end{align}
\end{subequations}
Note that the 8-derivative corrections decrease the mass {\it and} the entropy of an extremal black hole if $\lambda_e > 0$ and $\tilde{\lambda}_e > 0$.\footnote{This does not contradict our claim above because here we are comparing with an extremal Kerr black hole with the same $J$ as the corrected solution rather than a Kerr black hole with the same $M$ and $J$ as the corrected solution.}

This paper is organized as follows. In section \ref{general} we will prove our claim that the Euclidean action of a solution with higher derivative corrections is given, to first order, simply by evaluating the higher derivative terms on the Kerr solution. We will explain why odd-parity terms do not give first order corrections to thermodynamic quantities, and show that the 8-derivative terms contribute to the entropy (at fixed $M,J$) with a definite sign. In section \ref{Kerr} we will give explicit expressions for the first order corrections to thermodynamic quantities of the Kerr solution arising from ${\cal L}_6$ and ${\cal L}_8$. Finally section \ref{discussion} contains further discussion of our results.  

{\bf Conventions}. We work with a positive signature metric and units $\hbar =G_N=1$. Latin indices $a,b,\ldots$ are abstract indices, Greek indices $\mu,\nu,\ldots$ refer to a specific basis. Our curvature convention is $R^{\mu}{}_{\nu\rho\sigma} = \partial_\rho \Gamma^\mu_{\nu \sigma} + \ldots$. 

\section{General results}
\label{general}

In this section we will show that first order corrections to thermodynamic quantities can be calculated explicitly without determining the metric perturbation. The approach is based on the Euclidean action. 

We start by noting that higher derivative corrections do not affect the definition of the mass or angular momentum of the black hole in terms of surface integrals at spatial infinity: these are given by the same ADM or Komar formulae as in Einstein gravity. The temperature and angular velocity are also defined in the same way as in Einstein gravity. However, the higher derivative corrections do affect the definition of the entropy, which is now given by the Wald entropy \cite{Wald:1993nt}. This is the sum of the Bekenstein-Hawking entropy and a term coming from the higher derivatives in the action. 

The Euclidean action for Einstein gravity on a manifold $M$ with boundary $\partial M$ is \cite{hawking}
\be
I_{\rm Einstein} =-\frac{1}{16 \pi} \int_M d^4 x \sqrt{g} {\cal R}  - \frac{1}{8 \pi} \int_{\partial M} d^3 x \sqrt{h} K
\ee
where $h_{ab}$ is the metric induced on $\partial M$ and $K$ is the trace of the extrinsic curvature $K_{ab}$ of $\partial M$. We are interested in asymptotically flat black hole solutions, for which $\partial M$ has topology $S^1 \times S^2$ where the $S^1$ is the Euclidean time circle. The above expression is divergent as $\partial M$ approaches infinity (when the size of the $S^2$ diverges). To obtain a finite result we ``subtract the action of flat space" as follows  \cite{Gibbons:1976ue,hawking}. Take $\partial M$ to be a surface at finite radius $r=R$, where $r$ is defined by the property that the area of the $S^2$ is $4 \pi r^2$ for large $r$. We then define 
\be
 I_R[g] = I^R_{\rm Einstein}[g]-I^R_{\rm Einstein}[\delta_R] \, .
\ee
Here $I^R_{\rm Einstein}[g]$ is the Einstein action evaluated on the black hole metric, taking $M$ as the region $r \le R$, so $\partial M$ is the surface $r=R$, with induced metric $h_{ab}$. $I^R_{\rm Einstein}[\delta_R]$ is the Einstein action evaluated on a flat metric $\delta_R$ defined on $M' \equiv  S^1 \times \mathbb{R}^3$ with boundary $\partial M'$ diffeomorphic to $\partial M$ and induced metric $h'_{ab}$ on $\partial M'$. $\delta_R$ is chosen such that $h'_{ab}-h_{ab}$ vanishes as $R \rightarrow \infty$. For the Schwarzschild metric one can choose $h'_{ab} = h_{ab}$ \cite{Gibbons:1976ue}. For the Kerr metric, taking $r$ to be the Boyer-Lindquist radial coordinate, the metric differs from Schwarzschild only by terms of order $1/r^2$ (in a basis orthonormal w.r.t. the asymptotic flat metric) and so we can arrange that $h'_{\mu\nu}$ is the same as for the Schwarzschild metric with the same mass, which gives $h'_{\mu\nu} - h_{\mu\nu} = {\cal O}(1/R^2)$ (in an orthonormal basis). With this prescription, $I_R[g]$ has a limit as $R \rightarrow \infty$ so we define the renormalized action as
\be
 I_\infty[g] = \lim_{R \rightarrow \infty} I_R[g]
\ee
The result of this calculation for the Kerr metric is \cite{Gibbons:1976ue,hawking}
\be
\label{Kerr_action}
 I_\infty[{\rm Kerr}] = \frac{\beta \bar{M}}{2} 
\ee
where $\bar{M}$ is the ADM mass of the Kerr solution and $\beta = 1/T$ where $T$ is the Hawking temperature \cite{Hawking:1974sw}. 

We now include the higher-derivative terms. Since we are working to first order, we can calculate the effects of these terms by superposing the effects of the individual higher-derivative terms. The full action with one of these terms is
\be
 I = I_\infty[g] + \epsilon I_{\rm hd}[g]
\ee
where $\epsilon$ denotes one of the coefficients of the higher derivative term including any factors of $L$ (e.g. $\epsilon = \eta_e L^4$ or $\epsilon = \tilde{\lambda}_e L^6$) and 
\be
\label{Ihd}
 I_{\rm hd}[g] =-\frac{1}{16 \pi} \int_M d^4 x  \sqrt{g} {\cal L}_{\rm hd}  - \frac{1}{8 \pi} \int_{\partial M} d^3 x \sqrt{h} B
 \ee 
 where ${\cal L}_{hd}$ is the Lagrangian of the higher-derivative term stripped of coupling constants and factors of $L$ (e.g. ${\cal L} = {\cal C}^2$). The minus sign arises from rotation to Euclidean signature. $B$ denotes a possible higher-derivative boundary term which we will discuss below.
 
In Einstein gravity the Smarr relation \cite{Smarr:1972kt} implies that one can write \cite{Gibbons:1976ue}
\be
\label{IPhi}
 I = \beta G
\ee
where $G$ is the Gibbs free energy:
\be
G \equiv M - TS - \Omega_H J
\ee
where $M$ is the ADM mass, $S$ is the Bekenstein-Hawking entropy, $\Omega_H$ is the angular velocity and $J$ is the ADM angular momentum. 

Crucially for us, the result \eqref{IPhi} generalizes to a higher-derivative theory provided that we now interpret $S$ as the Wald entropy of the black hole and the boundary term $B$ of \eqref{Ihd} is chosen appropriately  \cite{Wald:1993nt}.
We regard $G$ as a function of $T$, $\Omega_H$ and $\epsilon$: from $G$ we can then determine the other thermodynamic quantities via the first law of black hole mechanics  \cite{Wald:1993nt} which gives $dG = -SdT - Jd\Omega_H$ (at fixed $\epsilon$), i.e., we have the standard relations:
\be
\label{Gtherm}
 S = - \left( \frac{\partial G}{\partial T} \right)_{\Omega_H,\epsilon} \qquad J = -\left( \frac{\partial G}{\partial \Omega_H} \right)_{T,\epsilon}
\ee
and then once we know $S,J$ we can calculate $M$ directly from the definition of $G$. So if we can determine the Euclidean action $I$ as a function of $(T,\Omega_H,\epsilon)$ then \eqref{IPhi} fixes all thermodynamic quantities of the black hole. 

We can calculate $I$ to first order in $\epsilon$ as follows. Let $g(T,\Omega_H,\epsilon)$ be the metric of our corrected Kerr solution, parameterized by temperature and angular velocity. Viewing $I$ as a function of $(T,\Omega_H,\epsilon)$ we have
\be
 I(T,\Omega_H,\epsilon) = I_\infty(T,\Omega_H) + \epsilon \left( \frac{\partial I}{\partial \epsilon} \right)_{T,\Omega_H}\Bigg|_{\epsilon=0} + {\cal O}(\epsilon^2)
\ee
where the first term is the Euclidean action of the Kerr solution in Einstein gravity (eq \eqref{Kerr_action}). The correction term $\partial I/\partial \epsilon$ involves two types of term: first from the fact that we are evaluating the Einstein terms in the action on the {\it corrected} metric and second from the higher derivative (i.e. $\epsilon$-dependent) terms in the action evaluated on the Kerr solution:
\be
\label{dIde}
 \left( \frac{\partial I}{\partial \epsilon} \right)_{T,\Omega_H}\Bigg|_{\epsilon=0}  =  \left( \frac{\partial I_\infty[g(T,\Omega_H,\epsilon)]}{\partial \epsilon} \right)_{T,\Omega_H}\Bigg|_{\epsilon=0} + I_{\rm hd}[\bar{g}(T,\Omega_H)]
\ee 
where $\bar{g}(T,\Omega_H) \equiv g(T,\Omega_H,0)$ is the Kerr metric. 

Naively, it appears that the first term above must vanish because we can view $g(T,\Omega_H,\epsilon)$ as a 1-parameter variation of the Kerr solution, and the Kerr solution extremizes the Einstein-Hilbert action so this term must vanish. However, this argument is a bit too quick. If we work at finite radius then it is true that the Kerr solution extremizes the Einstein-Hilbert action for a fixed boundary metric $h_{ab}$ on $\partial M$. But in our case we are not fixing the boundary metric on a finite radius surface. Instead we want to impose boundary conditions at infinity and calculate the action using the limiting procedure defined above. So we need to be more careful. Nevertheless, we will now show that the first term of \eqref{dIde} does indeed vanish. 

We use the result that, for metric variations around a solution of the Einstein equation, we have \cite{Brown:1992br}
\be
 \delta I_{\rm Einstein} = \frac{1}{16 \pi} \int_{\partial M} d^3 x \sqrt{h} \left( K^{ab} -K h^{ab} \right) \delta h_{ab}
\ee
This implies that for the regulated Einstein-Hilbert action $I_R$ we have
\be
\label{BY}
\left( \frac{\partial I_R[g(T,\Omega_H,\epsilon)]}{\partial \epsilon} \right)_{T,\Omega_H}\Bigg|_{\epsilon=0} = \frac{1}{16 \pi} \int_{\partial M} d^3 x \left[ \sqrt{h} \left( K^{ab} -K h^{ab} \right) \left( \frac{\partial h_{ab}}{\partial \epsilon} \right)_{T,\Omega_H}\Bigg|_{\epsilon=0} \right]
\ee
Here the square brackets indicate the difference between the enclosed quantity calculated for the Kerr background with boundary metric $h_{ab}$ and for the flat background $\delta_R$ with boundary metric $h'_{ab}$ defined as discussed above. (Note that $h'_{ab}$ depends on $\epsilon$ because in our procedure for regulating the Einstein action we want to choose the flat background metric so that $h'_{ab}$ is ``as close as possible" to $h_{ab}$, which depends on $\epsilon$.) Recall that $\partial M$ is the surface $r=R$ where $r$ is a suitable radial coordinate. We can choose this radial coordinate $r$ so that the corrected Euclidean Kerr metric takes the asymptotically flat form
\be
 g = \left( 1 - \frac{2M(T,\Omega_H,\epsilon)}{r} \right) d\tau_E^2 + \left( 1 + \frac{2M(T,\Omega_H,\epsilon)}{r} \right) dr^2 + r^2 \left( d\theta^2 + \sin^2 \theta d\phi^2 \right) + \ldots
\ee
where $\tau_E$ is the Euclidean time coordinate and the ellipsis denotes corrections that are ${\cal O}(1/r^2)$ w.r.t. the basis 
\be
 e_{\tau_E} = d\tau_E \qquad e_r = dr \qquad e_\theta = r d\theta \qquad e_\phi = r \sin \theta d\phi
\ee
which is orthonormal w.r.t. the flat metric $d\tau_E^2 + dr^2 + r^2 \left( d\theta^2 + \sin^2 \theta d\phi^2 \right)$. Note that the dependence on $\Omega_H$ enters from demanding that the Euclidean black hole metric is smooth when the coordinates are identified as $(\tau_E,\phi) \sim (\tau_E,\phi+2\pi) \sim (\tau_E + \beta, \phi +  i\Omega_H \beta)$.\footnote{
The Euclidean black hole metric is complex but the underlying manifold is real \cite{Brown:1990di}. Real coordinates can be defined by setting $\tilde{\phi} = \phi - i\Omega_H \tau_E$ so the identifications are $(\tau_E,\tilde{\phi}) \sim (\tau_E,\tilde{\phi}+2\pi) \sim (\tau_E + \beta, \tilde{\phi})$ \cite{Dias:2010eu}.} The induced metric on $\partial M$ is
\be
h = \left( 1 - \frac{2M(T,\Omega_H,\epsilon)}{R} \right) d\tau_E^2 + R^2 \left( d\theta^2 + \sin^2 \theta d\phi^2 \right) + \ldots
\ee
where the ellipsis denotes a term whose components are ${\cal O}(1/R^2)$ w.r.t. the basis ${\cal B} = \{e_{\tau_E},e_\theta,e_\phi\}$ (evaluated at $r=R$).

We now define the flat background metric as 
\be
 \delta_R = \left( 1 - \frac{2M(T,\Omega_H,\epsilon)}{R} \right) d\tau_E^2 + dr^2 + r^2 \left( d\theta^2 + \sin^2 \theta d\phi^2 \right) \ee
which gives the induced metric on the surface $r=R$ as
\be
 h' = \left( 1 - \frac{2M(T,\Omega_H,\epsilon)}{R} \right) d\tau_E^2 + R^2 \left( d\theta^2 + \sin^2 \theta d\phi^2 \right) 
\ee
This choice of $\delta_R$ ensures that $h_{\mu\nu} - h'_{\mu\nu}={\cal O}(1/R^2)$ in the basis ${\cal B}$. From these results we have, in this basis
\be
\label{hhp}
  \left( \frac{\partial h'_{\mu\nu}}{\partial \epsilon} \right)_{T,\Omega_H}\Bigg|_{\epsilon=0} = {\cal O}(1/R)   \; , \qquad\left( \frac{\partial h'_{\mu\nu}}{\partial \epsilon} \right)_{T,\Omega_H}\Bigg|_{\epsilon=0}-\left( \frac{\partial h_{\mu\nu}}{\partial \epsilon} \right)_{T,\Omega_H}\Bigg|_{\epsilon=0} = {\cal O}(1/R^2)
\ee
In this basis, the leading ${\cal O}(1/R)$ behaviour of the quantity $K^{\mu\nu} - K h^{\mu\nu}$ is the same for both the flat metric and the corrected Kerr metric (the mass appears only at subleading order). Equation \eqref{hhp} now implies that the leading ${\cal O}(1/R^2)$ behaviour of the scalar $(K^{\mu\nu} - K h^{\mu\nu})(\partial h_{\mu\nu} /\partial \epsilon)_{T,\Omega_H}|_{\epsilon=0}$ is the same for the corrected Kerr metric and for the flat background. Furthermore, the leading ${\cal O}(R^2)$ behaviour of the volume form on $\partial M$  is the same for both spacetimes. Hence the contributions from the corrected Kerr metric and the flat background metric to the RHS of \eqref{BY} tend to the {\it same} finite limit as $R \rightarrow \infty$ and so the RHS tends to zero when we evaluate the difference of the terms in the square brackets. 

We have shown that the first term on the RHS of (\ref{dIde}) vanishes hence, using \eqref{Kerr_action}, we have
\be
\label{Ihd2}
 I(T,\Omega_H,\epsilon) =\frac{\beta}{2} \bar{M}(T,\Omega_H)+ \epsilon  I_{\rm hd}[\bar{g}(T,\Omega_H)]+ {\cal O}(\epsilon^2)
\ee
where $\bar{M}(T,\Omega_H)$ is the ADM mass of the Kerr solution. Hence the first order correction to the Euclidean action is given simply by evaluating the higher derivative terms in the action on the unperturbed Kerr solution.\footnote{\label{ads}A similar result was observed in a calculation of higher derivative corrections to the Euclidean action of anti-de Sitter black branes \cite{Gubser:1998nz} (see also \cite{Pawelczyk:1998pb}) and black holes  \cite{Caldarelli:1999ar}. In the AdS case one regularizes the action by adding appropriate boundary terms \cite{Balasubramanian:1999re}. The analogue of \eqref{BY} is then an integral over the conformal boundary of AdS of $T^{ab}(\partial h_{ab}/\partial \epsilon)_T$ where $T^{ab}$ is the holographic energy-momentum tensor \cite{Balasubramanian:1999re} and $h_{ab}$ the metric on the conformal boundary. This integral vanishes because standard boundary conditions (and fixed $T$) fix $h_{ab}$ and therefore imply $(\partial h_{ab}/\partial \epsilon)_T=0$. This explains the observation of \cite{Gubser:1998nz}.} To calculate $I_{\rm hd}$ we only need to evaluate the bulk term in \eqref{Ihd}. This is because  $L_{\rm hd}$ is at least cubic in curvature, which implies that $B$ will decay rapidly as $\partial M$ is taken to infinity in an asymptotically flat spacetime. Hence $B$ will not contribute to the Euclidean action in the infinite volume limit. 

In summary, equations \eqref{IPhi} and \eqref{Ihd2} can be used to calculate thermodynamic quantities as functions of $(T,\Omega_H)$ to first order in $\epsilon$ without having to calculate the corrected metric. It is now immediate that {\it the odd parity terms do not affect thermodynamic quantities} (to first order in $\epsilon$). This is because the Kerr solution has an orientation reversing isometry $\theta \rightarrow \pi - \theta$ (in Boyer-Lindquist coordinates) which implies that the action of an odd parity term vanishes when evaluated on the Kerr solution. So henceforth we focus on the even parity terms.

Next we adapt an argument of \cite{Cheung:2018cwt} to relate the change in $S$ at fixed $(M,J)$ to the change in the action that we have just discussed. First from the definition of $G$ we have
\be
 \left[ \frac{\partial (\beta G)}{\partial \epsilon} \right]_{T,\Omega_H} = \beta \left( \frac{\partial M}{\partial \epsilon} \right)_{T,\Omega_H} - \beta \Omega_H \left( \frac{\partial J}{\partial \epsilon} \right)_{T,\Omega_H} - \left( \frac{\partial S}{\partial \epsilon} \right)_{T,\Omega_H}
\ee
The last term is
\be
\left( \frac{\partial S}{\partial \epsilon} \right)_{T,\Omega_H} = \left( \frac{\partial S}{\partial M} \right)_{J,\epsilon} \left( \frac{\partial M}{\partial \epsilon} \right)_{T,\Omega_H}
+\left( \frac{\partial S}{\partial J} \right)_{M,\epsilon} \left( \frac{\partial J}{\partial \epsilon} \right)_{T,\Omega_H} + \left( \frac{\partial S}{\partial \epsilon} \right)_{M,J} 
\ee
and the first law in the form $dM = TdS + \Omega_H dJ$ (at fixed $\epsilon$) gives $\beta = (\partial S/\partial M)_{J,\epsilon}$ and $\beta \Omega_H  = - (\partial S/\partial J)_{M,\epsilon}$. Combining these results we obtain
\be
 \left[ \frac{\partial (\beta G)}{\partial \epsilon} \right]_{T,\Omega_H} = -\left( \frac{\partial S}{\partial \epsilon} \right)_{M,J}  
\ee 
But $\beta G$ is the Euclidean action so evaluating at $\epsilon=0$ and multiplying by $\epsilon$ gives
\be
\epsilon \left( \frac{\partial S}{\partial \epsilon} \right)_{M,J} \Bigg|_{\epsilon=0} = -\epsilon I_{\rm hd}[\bar{g}(T,\Omega_H)]
\ee
We can choose to label the Kerr solution with $(M,J)$ instead of $(T,\Omega_H)$ and so this shows that, to first order in $\epsilon$, the change in the entropy at fixed $M,J$ is minus the Euclidean action of the higher derivative terms evaluated on the Kerr solution with the same $M,J$. 

This is interesting because, for the even parity 8-derivative terms, the RHS has a definite sign since the Lagrangian of each of these terms is the square of something.\footnote{
Actually this is a bit too quick because we are evaluating the action on a {\it complex} metric. But the scalars ${\cal C}$ and $\tilde{\cal C}$ depend only on $(r,\theta)$ and so they remain real when analytically continued to Euclidean signature. Thus the integrand is indeed the square of a real quantity.} Thus we see immediately that the sign of the correction to the entropy from an even parity 8-derivative term is equal to the sign of the coefficient $\lambda_e$ or $\tilde{\lambda}_e$ in the action (the minus sign above is cancelled by the minus sign in \eqref{Ihd} coming from the rotation to Euclidean signature). So an even parity 8-derivative term increases the entropy at fixed $M,J$ (and decreases $G$ at fixed $T,\Omega_H$) if, and only if, the coefficient of this term is positive. As noted in the Introduction, this is exactly the same condition as arises from demanding unitarity and analyticity of graviton scattering amplitudes \cite{Bellazzini:2015cra} or absence of superluminal graviton propagation \cite{Gruzinov:2006ie}.

Finally we will derive some useful scaling relations. If ${\cal L}_{\rm hd}$ is a term involving $2p$ derivatives then $I_{\rm hd}$ has dimensions of $[{\rm length}]^{4-2p}$. This implies that, when evaluated on the Kerr solution, the result must take the form
\be
 I_{\rm hd}[\bar{g}(T,\Omega_H)] = \Omega_H^{2p-4} f(\tau)
\ee
for some function $f$ where we have defined the dimensionless quantity
\be
\label{taudef}
 \tau = \frac{2\pi T}{\Omega_H}
\ee
From our results above, the correction to the Gibbs free energy is\footnote{Henceforth all derivatives w.r.t. $\epsilon$ are understood to be evaluated at $\epsilon=0$.}
\be 
\left(  \frac{\partial G}{\partial \epsilon} \right)_{T,\Omega_H} = \Omega_H^{2p-3} g(\tau)
\ee 
where $g(\tau) = \tau f(\tau)/(2\pi)$. From \eqref{Gtherm} the corrections to the entropy and angular momentum at fixed $(T,\Omega_H)$ are
\be
 \left(  \frac{\partial S}{\partial \epsilon} \right)_{T,\Omega_H} = -2\pi \Omega_H^{2p-4} g'(\tau), \qquad  \left(  \frac{\partial J}{\partial \epsilon} \right)_{T,\Omega_H} = 2\pi T\Omega_H^{2p-5} g'(\tau) - (2p-3) \Omega_H^{2p-4} g(\tau)
\ee
The correction to the mass is now
\be
\left(  \frac{\partial M}{\partial \epsilon} \right)_{T,\Omega_H} = \left(  \frac{\partial G}{\partial \epsilon} \right)_{T,\Omega_H} + T \left(  \frac{\partial S}{\partial \epsilon} \right)_{T,\Omega_H}  + \Omega_H  \left(  \frac{\partial J}{\partial \epsilon} \right)_{T,\Omega_H} = (4-2p) \Omega_H^{2p-3} g(\tau)
\ee
Hence at fixed $(T,\Omega_H)$, the perturbation in $M$ is proportional to the perturbation in $G$:
\be
\left(  \frac{\partial M}{\partial \epsilon} \right)_{T,\Omega_H}  = (4-2p) \left(  \frac{\partial G}{\partial \epsilon} \right)_{T,\Omega_H} .
\label{eq:nice}
\ee
\section{Evaluating the corrections}

\label{Kerr}

We have shown above that the corrections to thermodynamic properties of the Kerr solution can all be obtained from the Euclidean action $I_{\rm hd}$ of the higher derivative terms evaluated on the Kerr solution. We will now evaluate this quantity for each of the different (even parity) terms. The calculation is straightforward using computer algebra so we will just sketch a few of the details. 

It is convenient to parameterize the Kerr solution using its horizon radius $r_+$ (in Boyer-Lindquist coordinate) and the spin parameter $a=J/M$. These are related to $T$ and $\Omega_H$ by
\begin{equation}
a=\frac{1}{2\Omega_H \sqrt{1+\tau^2} \left(\tau+\sqrt{1+\tau^2} \right)}\,,\quad \text{and}\quad r_+=\frac{1}{2\Omega_H \sqrt{1+\tau^2}}\,,
\end{equation}
where we defined $\tau$ in equation \eqref{taudef}. Here we are assuming $\Omega_H>0$. This is not a restriction: the orientation-reversing diffeomorphism $\phi \rightarrow -\phi$ reverses the sign of $\Omega_H$. Since we are considering only even parity terms, the action must be invariant under $\Omega_H \rightarrow - \Omega_H$. Thus $G$ must also be invariant under this transformation. It then follows that all thermodynamic quantities must have definite parity under $\Omega_H \rightarrow - \Omega_H$ at fixed $T$. 

To evaluate $I_{\rm hd}$ it is convenient to define a new coordinate $x = \cos \theta$. The integral over the coordinates $\tau_E$ and $\phi$ just gives a factor $2 \pi \beta$, leaving an integral over $(r,x)$. We found that the integration is easiest if we integrate \emph{first} in $x\in[-1,1]$ and only afterwards in $r\in[r_+,\infty)$, otherwise, the integrals quickly become unwieldy. 

In what follows it will also be useful to define dimensionless variations of several thermodynamic quantities at fixed angular velocity and temperature. We denote the variation of a thermodynamic quantity $A$ by
\begin{equation}
\bar{\delta} A \equiv \frac{A-\bar{A}}{\bar{A}}\,,
\end{equation}
where barred quantities indicate the Kerr quantities computed with the same angular velocity and temperature as the corrected solution. In terms of the notation of the previous section this is
\be
\bar{\delta} A \equiv \frac{\epsilon}{\bar{A}} \left( \frac{\partial A}{\partial \epsilon} \right)_{T,\Omega_H}
\ee
We can decompose our results into a sum of individual contributions from the different higher derivative terms. So we will write (recall that the odd parity terms do not contribute)
 \begin{equation}
\bar{\delta}A = \eta_e\,L^4\,\Omega_H^4\,\bar{\Delta}_{\eta_e}A+\lambda_e\,L^6\,\Omega_H^6\,\bar{\Delta}_{\lambda_e}A+\tilde{\lambda}_e\,L^6\,\Omega_H^6\,\bar{\Delta}_{\tilde{\lambda}_e}A\,.
\end{equation}
where we have extracted appropriate powers of $\Omega_H$ to make the quantities $\bar{\Delta}_i A$ dimensionless.

We begin with the corrections to the Gibbs free energy which directly follows from evaluating the Euclidean action on the Kerr solution. Decomposing $\bar{\delta} G$ as above we obtain
\begin{subequations}
\begin{align}
&\bar{\Delta}_{\eta_e}G=\frac{16}{7}\left[1+2 \tau ^2-7 \tau ^4+\tau\left(5-7 \tau ^2\right) \sqrt{1+\tau ^2}  \right]\,,
\\
&\bar{\Delta}_{\lambda_e}G=-\frac{16}{5}\left(\tau +\sqrt{1+\tau ^2}\right)^2\Bigg[608+6025 \tau ^2+21580 \tau ^4+37140 \tau ^6+34230 \tau ^8+16275 \tau ^{10}\nonumber
\\&+3150 \tau ^{12}-1575 \tau  \left(1+\tau ^2\right)^5 \left(1+2 \tau ^2\right) \mathrm{arccot}\,\tau\Bigg]\,,
\\
&\bar{\Delta}_{\tilde{\lambda}_e}G=-\frac{64}{5}\left(\tau +\sqrt{1+\tau ^2}\right)^2\Bigg[592+6185 \tau ^2+21500 \tau ^4+37140 \tau ^6+34230 \tau ^8+16275 \tau ^{10}\nonumber
\\&+3150 \tau ^{12}-1575 \tau  \left(1+\tau ^2\right)^5 \left(1+2 \tau ^2\right) \mathrm{arccot}\,\tau\Bigg]\,.
\end{align}
\end{subequations}
Here $\mathrm{arccot} \, \tau$, with $\tau>0$, is defined to have range $(0,\pi/2)$. In deriving these expressions, as well as the ones we present below, we have assumed $\Omega_H>0$ which implies $\tau>0$. As noted above, all thermodynamic quantities have definite parity under $\Omega_H \rightarrow - \Omega_H$ (at fixed $T$) which implies that all of our $\bar{\Delta} A$ quantities must be {\it even} under $\tau \rightarrow - \tau$. Thus results for negative $\tau$ can be obtained by replacing $\tau$ with $|\tau|$ in our expressions. 

The perturbation to the mass is determined using \eqref{eq:nice}. Using $\bar{G} = \bar{M}/2$, this gives
\be
\label{DeltaM}
 \bar{\Delta}_{\eta_e} M = - \bar{\Delta}_{\eta_e} G \qquad \bar{\Delta}_{\lambda_e}M = -2 \bar{\Delta}_{\lambda_e}G \qquad \bar{\Delta}_{\tilde{\lambda}_e} M = -2 \bar{\Delta}_{\tilde{\lambda}_e}G
\ee
The entropy and angular momentum are obtained using \eqref{Gtherm}. This gives
\begin{subequations}
\begin{align}
&\bar{\Delta}_{\eta_e} S=-\frac{16}{7}\left[4-13 \tau ^2-21 \tau ^4-\tau \left(1+21 \tau ^2\right)  \sqrt{1+\tau ^2}\right]\,,
\\
&\bar{\Delta}_{\lambda_e} S=\frac{16}{5}\left(\tau +\sqrt{1+\tau ^2}\right)^2\Bigg\{608+6025 \tau ^2+21580 \tau ^4+37140 \tau ^6+34230 \tau ^8+16275 \tau ^{10} \nn
\\
&+3150 \tau ^{12}+5 \tau  \sqrt{1+\tau ^2} \left(2725+19154 \tau ^2+48978 \tau ^4+59808 \tau ^6+35385 \tau ^8+8190 \tau^{10}\right) \nn
\\
&-1575 \left(1+\tau ^2\right)^4 \left[\tau +3 \tau ^3+2 \tau ^5+\sqrt{1+\tau ^2} \left(1+17 \tau ^2+26 \tau ^4\right)\right] \mathrm{arccot}\,\tau\Bigg\}\,,
\\
&\bar{\Delta}_{\tilde{\lambda}_e} S=\frac{64}{5}\left(\tau +\sqrt{1+\tau ^2}\right)^2\Bigg\{592+6185 \tau ^2+21500 \tau ^4+37140 \tau ^6+34230 \tau ^8+16275 \tau ^{10}
\nn \\
&+3150 \tau ^{12}+5 \tau  \sqrt{1+\tau ^2} \left(2789+19090 \tau ^2+48978 \tau ^4+59808 \tau ^6+35385 \tau ^8+8190 \tau^{10}\right)
\nn \\
&-1575 \left(1+\tau ^2\right)^4 \left[\tau +3 \tau ^3+2 \tau ^5+\sqrt{1+\tau ^2} \left(1+17 \tau ^2+26 \tau ^4\right)\right]  \mathrm{arccot}\,\tau\Bigg\}\,,
\end{align}
\label{eq:crazy_S}
\end{subequations}
while for the angular momentum we find
\begin{subequations}
\begin{align}
&\bar{\Delta}_{\eta_e} J=-\frac{16}{7}\eta_e\left[3+21 \tau ^2+14 \tau ^4+14 \tau  \left(1+\tau ^2\right)^{3/2}\right]\,,
\\
&\bar{\Delta}_{\lambda_e} J=-\frac{16}{5}\left(\tau +\sqrt{1+\tau ^2}\right)^3 \Bigg\{608 \tau +6025 \tau ^3+21580 \tau ^5+37140 \tau ^7+34230 \tau ^9+16275 \tau ^{11}
\nn \\
&+3150 \tau ^{13}-\sqrt{1+\tau ^2} \Big(3040+16500 \tau^2+12130 \tau ^4-59190 \tau ^6-127890 \tau ^8-95550 \tau ^{10}
\nn \\
&-25200 \tau ^{12}\Big)-1575 \tau  \left(1+\tau ^2\right)^4 \left[\tau +3 \tau ^3+2 \tau ^5-2\sqrt{1+\tau ^2} \left(2-\tau^2-8 \tau^4\right)\right] \mathrm{arccot}\,\tau\Bigg\}\,,
\\
&\bar{\Delta}_{\tilde{\lambda}_e} J=-\frac{64}{5} \left(\tau +\sqrt{1+\tau ^2}\right)^3 \Bigg\{592 \tau +6185 \tau ^3+21500 \tau ^5+37140 \tau ^7+34230 \tau ^9+16275 \tau^{11}
\nn \\
&+3150 \tau ^{13}-\sqrt{1+\tau ^2} \Big(2960+16980 \tau^2+12050 \tau ^4-59190 \tau ^6-127890 \tau ^8-95550 \tau ^{10}
\nn \\
&-25200 \tau ^{12}\Big)-1575 \tau  \left(1+\tau ^2\right)^4 \left[\tau +3 \tau ^3+2 \tau ^5-2\sqrt{1+\tau ^2} \left(2-\tau^2-8 \tau^4\right)\right] \mathrm{arccot}\,\tau\Bigg\}\,.
\end{align}
\label{eq:crazy_J}
\end{subequations}
All quantities listed above remain finite in the extremal limit $\tau\to0$. In particular, for the angular momentum we find
\begin{equation}
\left.\bar{\delta}J\right|_{T=0}=-\frac{48}{7}\,L^4\Omega_H^4\,\eta_e+9728\,L^6\,\Omega_H^6\,\lambda_e+37888\,L^6\,\Omega_H^6\,\tilde{\lambda}_e\,,
\end{equation}
which upon inversion to first order in $\{\eta_e\,,\lambda_e\,,\tilde{\lambda}_e\}$ yields the result quoted in Eq.~(\ref{eq:angext}). Substituting this into the $\tau \rightarrow 0$ limits of $\bar{\delta} M$ and $\bar{\delta} S$ then gives eqs~(\ref{eq:Mext}) and ~(\ref{eq:Sext}). 

The Schwarzschild limit is obtained by taking $\Omega_H \rightarrow 0$ with $\Omega_H \tau$ fixed. This gives
\begin{subequations}
\begin{align}
&\left.\bar{\delta} G\right|_{\Omega_H=0} = -512 \pi ^4 L^4 T^4 \eta _e-65536 \pi ^6 L^6 T^6 \lambda _e\,,
\\
&\left.\bar{\delta}S\right|_{\Omega_H=0}=1536 \pi ^4 L^4 T^4 \eta _e+327680 \pi ^6 L^6 T^6 \lambda _e\,,
\\
&\left.\bar{\delta}J\right|_{\Omega_H=0}=-1024 \pi ^4 L^4 T^4 \eta _e+\frac{2031616}{11} \pi ^6 L^6 T^6 \lambda _e+\frac{28311552}{11} \pi ^6 L^6 T^6 \tilde{\lambda }_e\,.
\end{align}
\end{subequations}
Note that in this limit $\bar{\Delta}_{\tilde{\lambda}_e} G=\bar{\Delta}_{\tilde{\lambda}_e} S=0$. 

At first sight it is surprising that $\bar{\delta}J \ne 0$ in the Schwarzschild limit but recall that $\bar{\delta} J$ is the {\it relative} change in $J$. The absolute change in $J$ is $\bar{J} \bar{\delta} J$ which vanishes in the Schwarschild limit. In fact the expression for $\bar{\delta} J$ gives the perturbation in the moment of inertia, defined as ${\cal I} = J/\Omega_H$. A simple calculation shows that $\bar{\delta}{\cal I} = \bar{\delta} J$. 

\begin{figure}[h]
\centering
\includegraphics[width=0.9\linewidth]{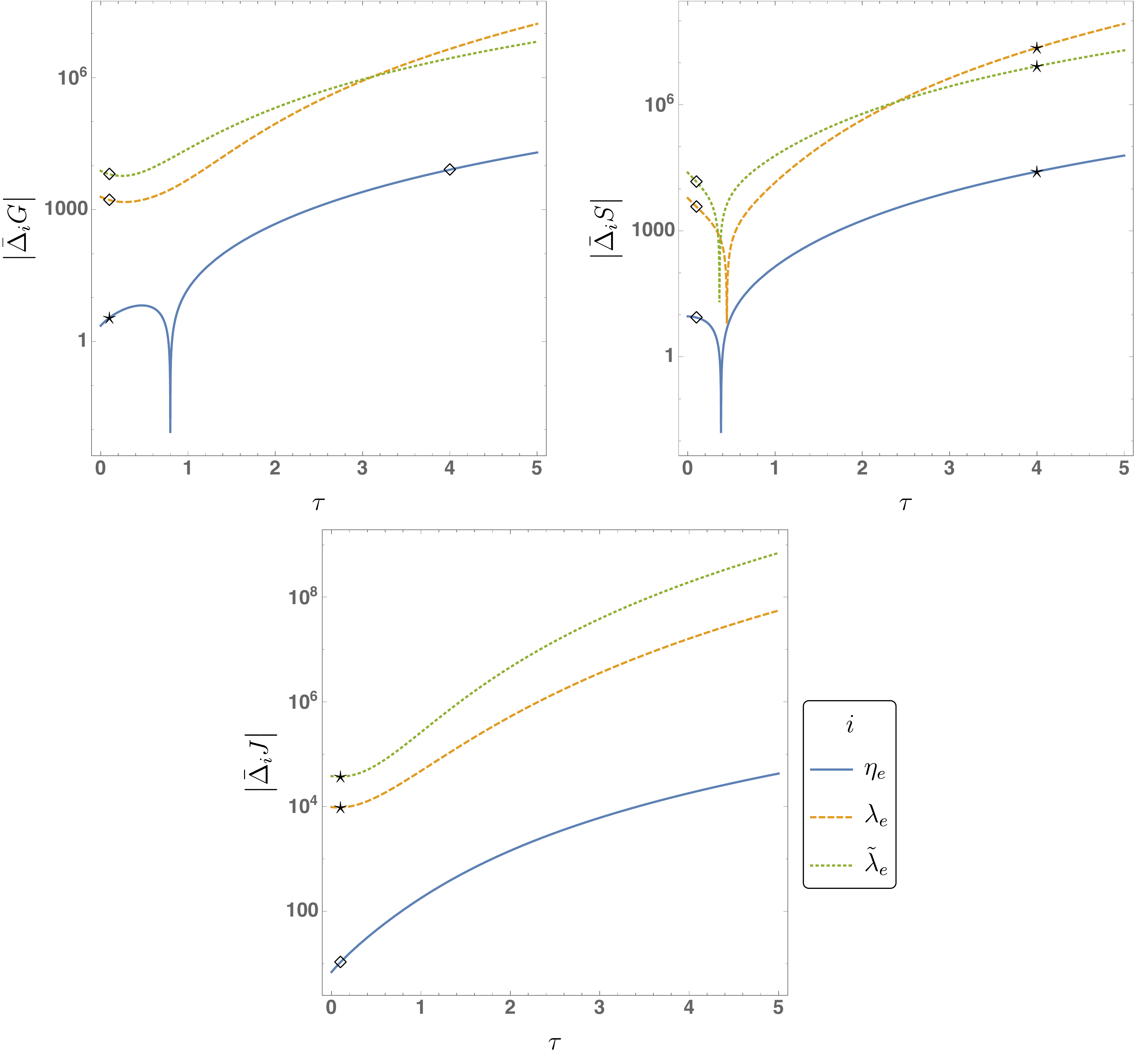}
\caption{Higher derivative corrections to Kerr thermodynamic quantities in the grand canonical ensemble. The extremal limit is $\tau \rightarrow 0$ and the Schwarzschild limit is $\tau \rightarrow \infty$. The stars and diamonds indicate whether, at that point, the quantity inside the modulus sign is positive or negative, respectively. The blue solid curves represent the coefficients of $\eta_e\,\Omega_H^4\,L^4$, the dashed orange curves indicate the coefficients of $\lambda_e L^6\Omega_H^6$ and the dotted green curves represent coefficients of $\tilde{\lambda}_e L^6\Omega_H^6$.}
\label{fig:1}
\end{figure}

In Fig.~\ref{fig:1} we plot, as a function of $\tau$, all perturbations of the relevant grand canonical thermodynamic quantitites on a logarithmic scale. The stars and diamonds indicate whether the quantity is positive or negative, respectively. The fact that $\bar{\Delta}_i G$ is always negative for the 8-derivative corrections was explained above. But we see now that $\bar{\Delta}_i J$ also has a definite sign, for both the 6-derivative and 8-derivative corrections. However, other quantities change sign as we vary $\tau$, in particular they might start positive for perturbations of the Schwarzschild black hole ($\tau\to+\infty$) and become negative at extremality, or \emph{vice-versa}. 

We now turn our attention to the microcanonical ensemble, where  $M$ and $J$ are held fixed. We can determine all of the above quantities in this ensemble, by inverting Eqs.~(\ref{DeltaM}) and~(\ref{eq:crazy_J}) to first order in $\{\eta_e\,,\lambda_e\,,\tilde{\lambda}_e\}$, to determine $\Omega_H$ and $T$ as a function of $M$ and $J$. It is natural to define the dimensionless quantity
\begin{equation}
j \equiv \frac{J}{M^2}\,.
\end{equation}
We introduce the following notation for the variations at fixed $M$ and $J$ of a thermodynamic quantity $A$:
\begin{equation}
\hat{\delta} A\equiv \frac{A-\hat{A}}{\hat{A}}\,,
\end{equation}
where hatted quantities indicate the Kerr quantities computed with the same energy and angular momentum as the corrected solution. We decompose $\hat{\delta} A$ as 
\begin{equation}
\hat{\delta}A = \eta_e\frac{L^4}{M^4}\hat{\Delta}_{\eta_e}A+\lambda_e\frac{L^6}{M^6}\hat{\Delta}_{\lambda_e}A+\tilde{\lambda}_e\frac{L^6}{M^6}\hat{\Delta}_{\tilde{\lambda}_e}A\,.
\end{equation}
After some lengthy algebra, one gets the following rather cumbersome result for the corrections to the angular velocity
\begin{subequations}
\begin{align}
&\hat{\Delta}_{\eta_e}\Omega_H=\frac{1}{7}\frac{1}{\left(1-j^2\right) \left(1+\sqrt{1-j^2}\right)^3}\left[18-34 j^2+16 j^4+\sqrt{1-j^2} \left(17-24 j^2+8 j^4\right)\right]\label{eq:angmatch}\,,
\\
&\hat{\Delta}_{\lambda_e}\Omega_H=\frac{2}{5}\frac{1}{ j^{10} \sqrt{1-j^2}\left(\sqrt{1+j}+\sqrt{1-j}\right)^6}\Bigg\{31500-54600 j^2+23100 j^4-720 j^6+160 j^8\nonumber
\\
&+80 j^{10}-384 j^{12}+256 j^{14}-\sqrt{1-j^2} \Big(3150-2625 j^2+105 j^4-30 j^6+40 j^8-160 j^{10}\nonumber
\\
&+128 j^{12}\Big)+1575\left(1-j^2\right)\left[2-j^2-4 \left(5-2 j^2\right) \sqrt{1-j^2}\right]\frac{\arcsin j}{j}\Bigg\}\,,
\\
&\hat{\Delta}_{\tilde{\lambda}_e}\Omega_H=\frac{8}{5}\frac{1}{j^{10} \sqrt{1-j^2}\left(\sqrt{1+j}+\sqrt{1-j}\right)^6}\Bigg\{31500-54600 j^2+23100 j^4-720 j^6+80 j^8\nonumber
\\
&-80 j^{10}+384 j^{12}-256 j^{14}-\sqrt{1-j^2} \Big(3150-2625 j^2+105 j^4-30 j^6-40 j^8+160 j^{10}\nonumber
\\
&-128 j^{12}\Big)+1575\left(1-j^2\right)\left[2-j^2-4 \left(5-2 j^2\right)\sqrt{1-j^2}\right]\frac{\arcsin j}{j}\Bigg\}\,,
\end{align}
\end{subequations}
where $\arcsin j \in (-\pi/2,\pi/2)$. For the temperature correction we find
\begin{subequations}
\begin{align}
&\hat{\Delta}_{\eta_e}T=\frac{1}{7}\frac{1}{\left(1-j^2\right) \left(1+\sqrt{1-j^2}\right)^3}\left[5-22 j^2+16 j^4+2 \sqrt{1-j^2} \left(1-6 j^2+4 j^4\right)\right]\label{eq:tempmatch}\,,
\\
&\hat{\Delta}_{\lambda_e}T=-\frac{2}{5}\frac{1}{j^{10} \left(1-j^2\right)\left(\sqrt{1+j}+\sqrt{1-j}\right)^6}\Bigg\{40950-71925 j^2+31290 j^4-1125 j^6+\nonumber
\\
&370 j^8-520 j^{10}+480 j^{12}-128 j^{14}-\sqrt{1-j^2} \Big(40950-62475 j^2+23415 j^4-810 j^6+280 j^8\nonumber
\\
&-400 j^{10}+256 j^{14}\Big)+1575\left(1-j^2\right)\left[26-31 j^2+8 j^4-\sqrt{1-j^2} \left(26-11 j^2\right)\right]\frac{\arcsin j}{j}\Bigg\}\,,
\\
&\hat{\Delta}_{\tilde{\lambda}_e}T=-\frac{8}{5}\frac{1}{j^{10} \left(1-j^2\right)\left(\sqrt{1+j}+\sqrt{1-j}\right)^6}\Bigg\{40950-71925 j^2+31290 j^4-1125 j^6\nonumber
\\
&+50 j^8+520 j^{10}-480 j^{12}+128 j^{14}-\sqrt{1-j^2} \Big(40950-62475 j^2+23415 j^4-810 j^6-40 j^8\nonumber
\\
&+400 j^{10}-256 j^{14}\Big)+1575\left(1-j^2\right)\left[26-31 j^2+8 j^4-\sqrt{1-j^2} \left(26-11 j^2\right)\right] \frac{\arcsin j}{j}\Bigg\}\,.
\end{align}
\end{subequations}
We have checked that expanding our expressions for $\hat{\Delta}_{\eta_e} \Omega_H$ and $\hat{\Delta}_{\eta_e} T$ in powers of $j$ reproduces the perturbative results of \cite{Cano:2019ore} for the 6-derivative terms. 

\begin{figure}[h]
\centering
\includegraphics[width=0.9\linewidth]{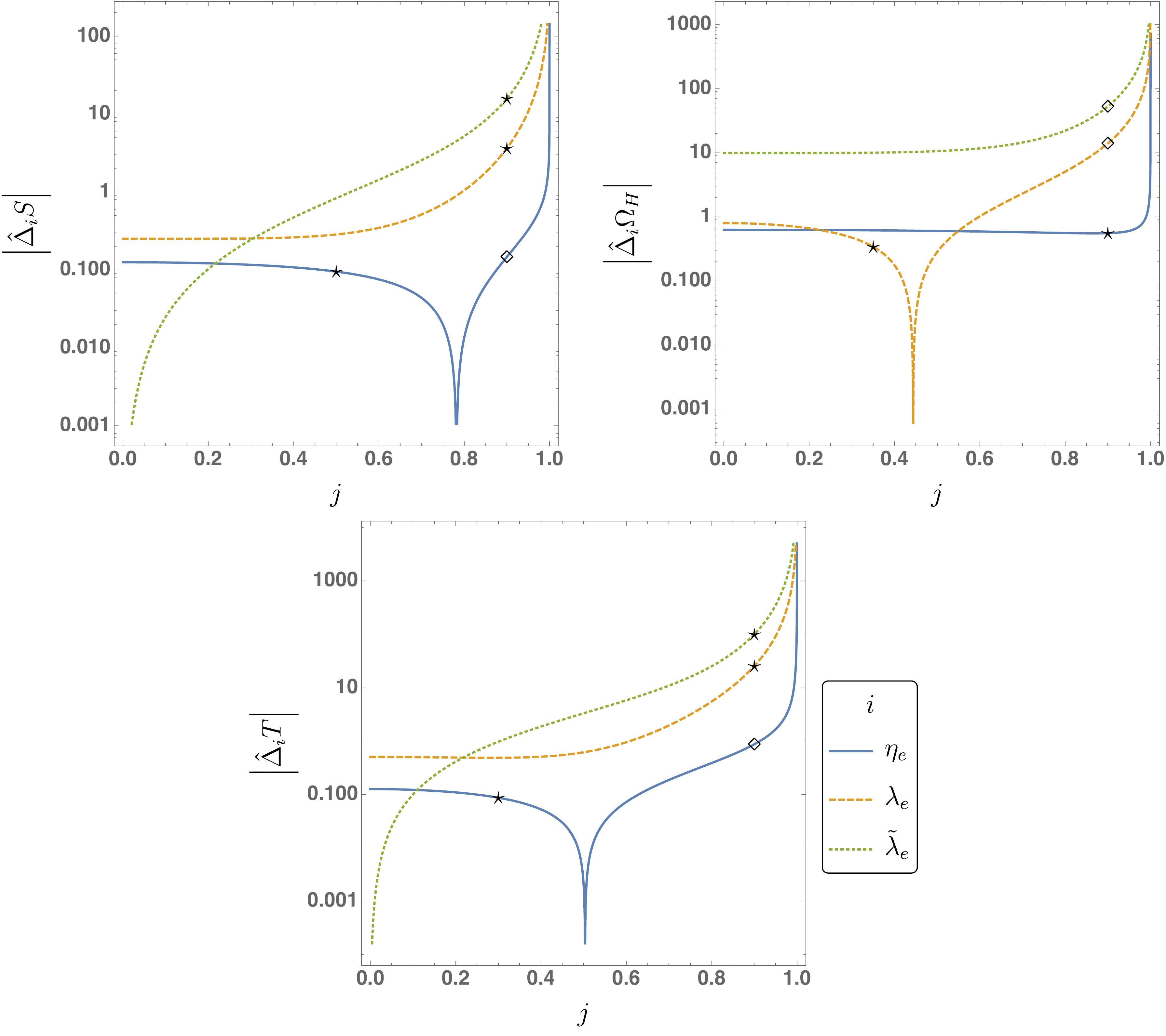}
\caption{Higher derivative corrections to Kerr thermodynamic quantities in the microcanonical ensemble. The Schwarzschild limit is $j \rightarrow 0$ and the extremal limit is $j \to 1$. The stars and diamonds indicate whether, at that point, the quantity inside the modulus sign is positive or negative, respectively. The blue solid curves represent terms proportional to $\eta_e L^4\Omega_H^4$, the dashed orange curves indicate terms proportional to $\lambda_e L^6\Omega_H^6$ and the dotted green curves represent terms proportional to $\tilde{\lambda}_e L^6\Omega_H^6$. The dotted green curves in the entropy and temperature plots vanish when $j=0$.}
\label{fig:2}
\end{figure}

Finally, we can write the entropy correction as a function of $M$ and $J$, which results in surprisingly compact expressions:
\begin{subequations}
\begin{align}
\hat{\Delta}_{\eta_e}S &=\frac{1}{7}\frac{1}{\sqrt{1-j^2}\left(1+\sqrt{1-j^2}\right)^3}\left[3-4 j^2+4 \left(1-2 j^2\right) \sqrt{1-j^2}\right]
\\
\hat{\Delta}_{\lambda_e}S&=\frac{1}{20}\frac{1}{j^8 \sqrt{1-j^2}\left(1+\sqrt{1-j^2}\right)^4}\Bigg[3150-2625 j^2+105 j^4-30 j^6+40 j^8-160 j^{10}\nonumber
\\
&+128 j^{12}-1575 \left(2-j^2\right)\sqrt{1-j^2}\;\frac{\arcsin j}{j}\Bigg]
\\
\hat{\Delta}_{\tilde{\lambda}_e}S&=\frac{1}{5}\frac{1}{j^8 \sqrt{1-j^2}\left(1+\sqrt{1-j^2}\right)^4}\Bigg[3150-2625 j^2+105 j^4-30 j^6-40 j^8+160 j^{10}\nonumber
\\
&-128 j^{12}-1575 \left(2-j^2\right)\sqrt{1-j^2}\;\frac{\arcsin j}{j}\Bigg]\,.
\end{align}
\label{eq:simple_S}
\end{subequations}
In Fig.~\ref{fig:2}, we plot the above quantities as functions of $j$. When $|j|\to 1$ all these quantities diverge. This divergence is not physical. It arises because $|j| \to 1$ corresponds to the extremal limit of the Kerr spacetime, but not to the extremal limit of the corrected solution. To see this, using the notation of section \ref{general}, let $M_{\rm ext}(J,\epsilon)$ denote the mass of the extremal corrected solution. Now consider, say, the entropy of the corrected solution with mass $\delta M$ above extremality, i.e., $S(M_{\rm ext}(J,\epsilon)+\delta M, J ,\epsilon)$. For fixed $\delta M$, the correction to the entropy of this solution is
\be
 \left( \frac{\partial S}{\partial \epsilon} \right)_{\delta M, J} \Bigg|_{\epsilon=0}= \left( \frac{\partial S}{\partial M} \right)_{J,\epsilon} \left( \frac{\partial M_{\rm ext}}{\partial \epsilon} \right)_J +  \left( \frac{\partial S}{\partial \epsilon} \right)_{M, J} = \frac{1}{T} \left( \frac{\partial M_{\rm ext}}{\partial \epsilon} \right)_J +  \left( \frac{\partial S}{\partial \epsilon} \right)_{M, J}
\ee
where we used the first law in the second equality and the RHS is evaluated at $M=M_{\rm ext}(J,0)+\delta M$ and $\epsilon=0$. Now take the extremal limit $\delta M \rightarrow 0$. If the corrected solution remains smooth in the extremal limit (as indicated by our grand canonical results) then the LHS approaches a finite limit. However, $T \rightarrow 0$ so if $(\partial M_{\rm ext}/\partial \epsilon)_{J} \ne 0$ then $(\partial S/\partial \epsilon)_{M,J}$ must diverge in order for the RHS to remain finite. Hence if the corrections change the mass of the extremal solution (as we have found) then $(\partial S/\partial \epsilon)_{M,J}$ must diverge in the extremal limit.\footnote{
Alternatively, consider the entropy $S(M,J,\epsilon)$ of the corrected solution. Obviously $S(M,J,0)$ is the entropy of the Kerr solution. This is analytic in $M,J$ {\it except} at extremality (because $(\partial S/\partial M)_J = 1/T$ diverges at extremality). Hence the function $S(M,J,\epsilon)$ is not analytic at $|j|=1$, $\epsilon=0$. Therefore it need not admit a Taylor expansion in $\epsilon$ when $|j|=1$. For example, the entropy might depend on a function like 
  $\sqrt{1-j^2+\alpha \epsilon}$ where $\alpha$ is a constant. This can be Taylor expanded in $\epsilon$ when $|j| <1$ but is proportional to $\sqrt{\epsilon}$ when $|j|=1$. The analogue of the corrected extremal limit for this example is $j^2 = 1+\alpha \epsilon$.} To study this limit it is better to work in the grand canonical ensemble where such a divergence does not occur because the limit is simply $T=0$, which is independent of $\epsilon$. 

Returning to Fig.~\ref{fig:2}, we see that the the entropy correction arising from the 8-derivative terms has a definite sign, as expected from the argument we gave in the previous section. However, the entropy correction arising from the 6-derivative term changes sign as the spin of the black hole increases. Similarly, the correction to the temperature arising from the 8-derivative terms has a definite sign (the same as that of the entropy correction) but the temperature correction arising from the 6-derivative term changes sign as the black hole spin increases. Finally, the angular velocity correction has a definite sign for the 6-derivative term and one of the 8-derivative terms but not for the other 8-derivative term. 


For the 8-derivative corrections, several of the above quantities appear to diverge as $j\to0$. However, these divergences cancel and we find
\begin{subequations}
\begin{align}
&\hat{\delta} S=\frac{L^4}{8 M^4}\eta _e+\frac{L^6}{4 M^6}\lambda _e\,,
\\
&\hat{\delta} T = \frac{L^4}{8 M^4}\eta _e+\frac{L^6 }{2 M^6}\lambda _e\,,
\\
&\hat{\delta} \Omega_H = \frac{5 L^4}{8 M^4} \eta _e+\frac{35 L^6}{44 M^6}\lambda _e-\frac{108 L^6}{11 M^6}\tilde{\lambda }_e\,.
\end{align}
\end{subequations}
Note that in this limit $\hat{\Delta}_{\tilde{\lambda}_e} S=\hat{\Delta}_{\tilde{\lambda}_e} T=0$. This is in agreement with our results for the grand canonical ensemble: the $\tilde{\lambda}_e$ term does not correct thermodynamic quantities of the Schwarzschild solution.\footnote{The explanation for why $\hat{\delta} \Omega_H$ is non-zero in the Schwarzschild limit is the same as the one for $\bar{\delta} J$ discussed above.}


\section{Discussion}

\label{discussion}

We have shown how to calculate, to first order, higher derivative corrections to thermodynamic quantities of the Kerr black hole {\it without determining the metric perturbation}. Clearly this method is not restricted to the Kerr solution and it is likely to have other applications. For example it can be applied to anti-de Sitter black holes (footnote \ref{ads}). It is likely to be particularly valuable for studying corrections to the thermodynamics of solutions which depend non-trivially on several coordinates, or solutions which have been obtained numerically. 

This method makes it clear that odd parity terms in the action do not lead to first order corrections to any thermodynamic quantities. However, the results of Refs. \cite{Cardoso:2018ptl,Cano:2019ore,reallsantos} show that these terms do give a non-vanishing correction to the metric. 

We have shown that the 8-derivative term with coefficient $\tilde{\lambda}_e$ gives vanishing first order corrections to thermodynamic quantities of the {\it Schwarzschild} black hole. However, the corrections due to this term are non-vanishing for a rotating Kerr black hole. 

The corrections to the entropy (at fixed $M$, $J$) arising from the 8-derivative terms are non-negative if, and only if, the corresponding coupling constants are non-negative. As noted in the Introduction, this is exactly the same condition that arises from unitarity and analyticity of graviton scattering amplitudes  \cite{Bellazzini:2015cra} and from absence of superluminal graviton propagation \cite{Gruzinov:2006ie}. Therefore it is interesting to ask whether there is some reason why these corrections to the entropy {\it had} to be positive. For example, if the higher derivative corrections arise from integrating out massive fields then one might expect the corrections to increase the entropy at fixed mass and angular momentum. This is because with extra massive degrees of freedom one might expect there to exist more black hole microstates with the given $M$ and $J$ than without these degrees of freedom (see e.g. Refs \cite{Goon:2016mil,Cheung:2018cwt}).
However, one has to be cautious about this argument. The degrees of freedom responsible for black hole entropy are presumably quantum gravity, i.e. UV, degrees of freedom. Consider a low energy theory with a massive field and the same theory without this field. The argument just presented assumes (i) that each of these low energy theories admits a UV completion, and (ii) that the UV completion of the former theory contains more degrees of freedom than the UV completion of the latter theory. Both assumptions are questionable.

The significance of our result about the sign of the 8-derivative entropy correction appears undermined by our results for the 6-derivative terms. We have seen that the sign of the entropy correction (at fixed $M,J$) coming from the even-parity ${\cal L}_6$ term changes sign as the spin is increased. So even if one demands that this correction is positive for a Schwarzschild black hole, it will become negative for a rapidly rotating Kerr hole (or vice versa). 

Ref. \cite{Cheung:2018cwt} gives a nice argument that the entropy correction should be positive if (a) the higher-derivative terms arise from integrating out a massive field at {\it tree-level} and (b) the black hole is thermodynamically stable in the sense of minimizing, rather than simply extremizing, the Euclidean action. This does not apply in our case because the Kerr black hole never satisfies (b) \cite{Monteiro:2009tc,Dias:2010eu}. However, a Kerr-AdS black hole {\it does} satisfy (b) if it is large enough and doesn't rotate too fast \cite{Monteiro:2009ke}. We have repeated our calculation for this case (replacing the Riemann tensor with the Weyl tensor in ${\cal L}_6$) and we find that the  ${\cal L}_6$ correction to the entropy of a thermodynamically stable black hole does not have a definite sign. So in the AdS case this suggests that the even-parity ${\cal L}_6$ term cannot be generated as in (a).

For the scenario with macroscopic $L$ discussed in \cite{Endlich:2017tqa}, the ${\cal L}_6$ terms are disfavoured on theoretical grounds because of arguments of \cite{Camanho:2014apa}. Specifically, if one wishes to avoid superluminal gravitons arising from these terms then one has to add extra degrees of freedom. If these eliminate the superluminality {\it at tree level} then Ref. \cite{Camanho:2014apa} argues that they must consist of a tower of higher spin fields with gravitational strength interactions with Standard Model fields. This is excluded observationally for macroscopic $L$  \cite{Endlich:2017tqa}. 

The arguments of the previous two paragraphs concern tree-level effects. But ${\cal L}_6$ can also be generated at 1-loop simply by integrating our a massive free field. In this case, the sign of $\eta_e$ is opposite for spin-0 and spin-1/2 fields, so these give entropy corrections of opposite sign \cite{Goon:2016mil}. So if one imposes some restriction on the sign of the entropy correction for, say, slowly rotating holes, then this would imply a constraint on the spectrum of massive fields of different types. And even with such a constraint, the entropy correction would change sign for rapidly rotating holes. 

Since we are now discussing a 1-loop effect arising from massive fields we should also consider the size of 1-loop effects arising from massless fields, including the gravitational field. These give a contribution to the entropy proportional to $\log A$ where $A\sim r_+^2$ is the horizon area in Planck units \cite{Solodukhin:1994yz,Solodukhin:1994st,Fursaev:1994te,Sen:2012dw}. This is much larger than the correction arising from integrating out a massive field at 1-loop. To see this, note that we can only integrate out a field of mass $m$ when $m r_+ \gg 1$ and the correction to the entropy from the resulting ${\cal L}_6$ is of order $1/(mr_+)^2$, which is small compared to the $\log A$ term arising from the massless fields. 

In the scenario of Ref. \cite{Endlich:2017tqa} the higher derivative corrections are much larger than the quantum contribution from the massless fields. For example, the 8-derivative correction to the entropy scales as $L^2 (L/r_+)^4$ where $r_+$ is the horizon radius. $L/r_+$ must be small for EFT to be trusted but this contribution to the entropy can still be large compared to $\log A \sim \log r_+$ if $L$ is much larger than the Planck length, as envisaged in \cite{Endlich:2017tqa}.

Finally, we have determined the corrections to the properties of {\it extremal} black holes. If the coefficients of the 8-derivative terms have the ``good" sign discussed above then the effect of these terms is to reduce the mass and entropy of an extremal black hole with fixed $J$. But the 6-derivative terms increase the mass and reduce the entropy, or vice-versa. One might wonder whether the higher-derivative correction to the metric remains smooth at the horizon in the extremal limit. It turns out that it does \cite{reallsantos}.

\section*{Acknowledgments}
We are grateful to Garrett Goon and Gary Horowitz for useful discussions. This work was supported by STFC grant No. ST/P000681/1.

\bibliography{higher_derivative}{}
\bibliographystyle{JHEP}
\end{document}